\documentclass[epj,twocolumn,fleqn]{webofc}
\usepackage[varg]{txfonts}   % Web of Conferences font
\usepackage{graphicx}

\newcommand{\ie}{\textit{i.e.}}

\newcommand{\vlp}{\left[\omega + m\Omega\right]}

\newcommand{\vect}{\vec}
\newcommand{\grad}{\vect{\nabla}}

\newcommand{\lapl}{\Delta}
\renewcommand{\div}{\vect{\nabla} \cdot}
\renewcommand{\l}{\ell}

\newcommand{\vFR}{\vect{F}^{\mathrm{R}}}

\newcommand{\Teff}{T_{\mathrm{eff}}}
\newcommand{\geff}{g_{\mathrm{eff}}}
\newcommand{\dpart}[2]{\frac{\partial #1}{\partial #2}}

\newcommand{\dTeff}{\delta T_{\mathrm{eff}}}
\newcommand{\dgeff}{\delta g_{\mathrm{eff}}}

\newcommand{\OmegaK}{\Omega_{\mathrm{K}}}

\wocname{epj}
\woctitle{Seismology of the Sun and the Distant Stars 2016}
\begin{document}
%***********************************************************************
\title{Non-adiabatic pulsations in ESTER models}

\author{\firstname{Daniel Roy} \lastname{Reese}\inst{1}\fnsep\thanks{\email{daniel.reese@obspm.fr}} \and
        \firstname{Marc-Antoine} \lastname{Dupret}\inst{2} \and
        \firstname{Michel} \lastname{Rieutord}\inst{3,4}
}

\institute{LESIA, Observatoire de Paris, PSL Research University, CNRS, Sorbonne Universités,
           UPMC Univ. Paris 06, Univ. Paris Diderot, Sorbonne Paris Cité, 92195 Meudon, France
\and
           Institut d'Astrophysique et Géophysique de l'Université de Liège, Allée du 6 août 17,
           4000 Liège, Belgium
\and
           Université de Toulouse, UPS-OMP, IRAP, 31028 Toulouse, France
\and
           CNRS, IRAP, 14 avenue Edouard Belin, 31400 Toulouse, France
          }

%-----------------------------------------------------------------------
\abstract{%
One of the greatest challenges in interpreting the pulsations of rapidly
rotating stars is mode identification, \ie\ correctly matching theoretical modes
to observed pulsation frequencies. Indeed, the latest observations as well as
current theoretical results show the complexity of pulsation spectra in such
stars, and the lack of easily recognisable patterns.  In the present
contribution, the latest results on non-adiabatic effects in such pulsations are
described, and we show how these come into play when identifying modes.  These
calculations fully take into account the effects of rapid rotation, including
centrifugal distortion, and are based on models from the ESTER project,
currently the only rapidly rotating models in which the energy conservation
equation is satisfied, a prerequisite for calculating non-adiabatic effects. 
Non-adiabatic effects determine which modes are excited and play a key role in
the near-surface pulsation-induced temperature variations which intervene in
multi-colour amplitude ratios and phase differences, as well as line profile
variations.
}
\maketitle
%
%-----------------------------------------------------------------------
\section{Introduction}
\label{intro}

The interpretation of pulsation spectra in rapidly rotating stars has proven to
be a considerable challenge.  Indeed, due to the lack of simple and easily
identifiable pattern in the pulsation spectra, it has been very difficult to
establish the correspondence between observed pulsation frequencies and
theoretically calculated modes, \ie\ to carry out mode identification
\cite{Goupil2005}. This is extremely problematic since it means we do not know
what regions of the star are being probed by individual modes, and cannot,
accordingly, carry out meaningful inversions of the stellar structure. 
Nonetheless, \cite{Bedding2015} and \cite{VanReeth2016} have recently shown the
presence of patterns in the period spacings of rapidly rotating $\gamma$ Dor
stars, which agree with the most recent theoretical expectations
\cite{Ballot2012, Bouabid2013, Ouazzani2016} and allow us, in some cases, to
determine the dominant harmonic degree and azimuthal order of these modes.  For
$\delta$ Scuti stars, progress remains more limited.  In a recent paper,
\cite{GarciaHernandez2015} derived an observational scaling relation between the
large frequency separation and the mean density of several $\delta$ Scuti stars
in eclipsing binary systems, in agreement with previous theoretical predictions
\cite{Reese2008a}.  Such large separations have also been used to create echelle
diagrams \cite{GarciaHernandez2013, Paparo2016}. However, no firm identification
has been obtained for individual modes.

As such, it is necessary to develop ways to observationally constrain mode 
identification, in addition to searching for patterns in the observed frequency
spectra.  In slowly rotating stars, amplitude ratios and phases differences
based on multi-colour photometry, as well as line profile variations observed
via spectroscopy, have proven to be popular methods for constraining mode
identification \cite{Dupret2003,Zima2006}. Various efforts have been made
to extend these methods to rapid rotators, but it is only recently that all of
the necessary ingredients are starting to come together. These include:
\begin{enumerate}
\item rapidly rotating models in which the energy conservation equation is
      satisfied. Indeed, this is important for carrying out self-consistent
      non-adiabatic pulsation calculations as will be described in the following
      point. Currently the only code able to produce such models is the
      ESTER\footnote{``Evolution STEllaire en Rotation''} project\footnote{We
      note that the Self-Consistent Field method only solves the energy
      conservation equation after the averaging the model over level surfaces.}
      \cite{Rieutord2013, EspinosaLara2013, Rieutord2016}.
\item non-adiabatic pulsation calculations which fully take into account the 
      effects of rapid rotation.  Indeed, these can be used to predict which
      modes are excited by the $\kappa$-mechanism, the predominant
      mode-excitation mechanism in rapid rotators.  Furthermore, only such
      calculations provide realistic pulsation-induced surface effective
      temperature variations, a key ingredient in photometric and spectroscopic
      mode signatures \cite{Dupret2002, Dupret2003}.   Up until recently, the
      only non-adiabatic code which fully takes into account the effects of
      rotation was that of \cite{Lee1995}.  However, this code was generally
      applied with a limited number of spherical harmonics to models based on a
      Chandrasekhar expansion, or even spherical models.  This has,  however,
      changed with the advent of the non-adiabatic version of the TOP code
      \cite{Reese2006, Reese2009a}, currently the only code which is set up to
      calculate pulsation modes in ESTER models.
\item detailed calculations of photometric mode visibilities and spectroscopic
      line profile variations (LPVs). Recently, \cite{Reese2013} carried out
      detailed mode visibility calculations in rapidly rotating SCF models
      \cite{MacGregor2007}.  However these calculations, now need to be extended
      to the non-adiabatic case as described above. Probably the most realistic
      LPV calculations in rapidly rotating stars are those of
      \cite{Townsend1997}, but were nonetheless carried out using the
      traditional approximation.  Other calculations \cite{Clement1994} fully
      include the effects of rotation on pulsations but use a simplified
      modelling of spectroscopic effects.
\end{enumerate}

In the following, we describe the latest non-adiabatic pulsation calculations in
ESTER models using the TOP code (Section~\ref{sect:theory}), and some of the
results for mode-excitation, mode visibilities, and line profile variations
(Section~\ref{sect:results}).

%-----------------------------------------------------------------------
\section{Theory}
\label{sect:theory}

\subsection{ESTER models}

The results presented in the following sections are based on 9 M$_{\odot}$ ESTER
models with rotation rates ranging from $0.0$ to $0.8$ $\OmegaK$, where
$\OmegaK$ is the Keplerian break-up rotation rate. These models
have a metallicity of $Z=0.025$ and use OPAL opacities.  They harbour $\beta$
Cep type pulsations, in the p and g-mode range, which are excited by the iron
opacity bump at $\log(T)=5.3$.

\begin{figure}[h]
\centering
\includegraphics[width=\hsize,clip]{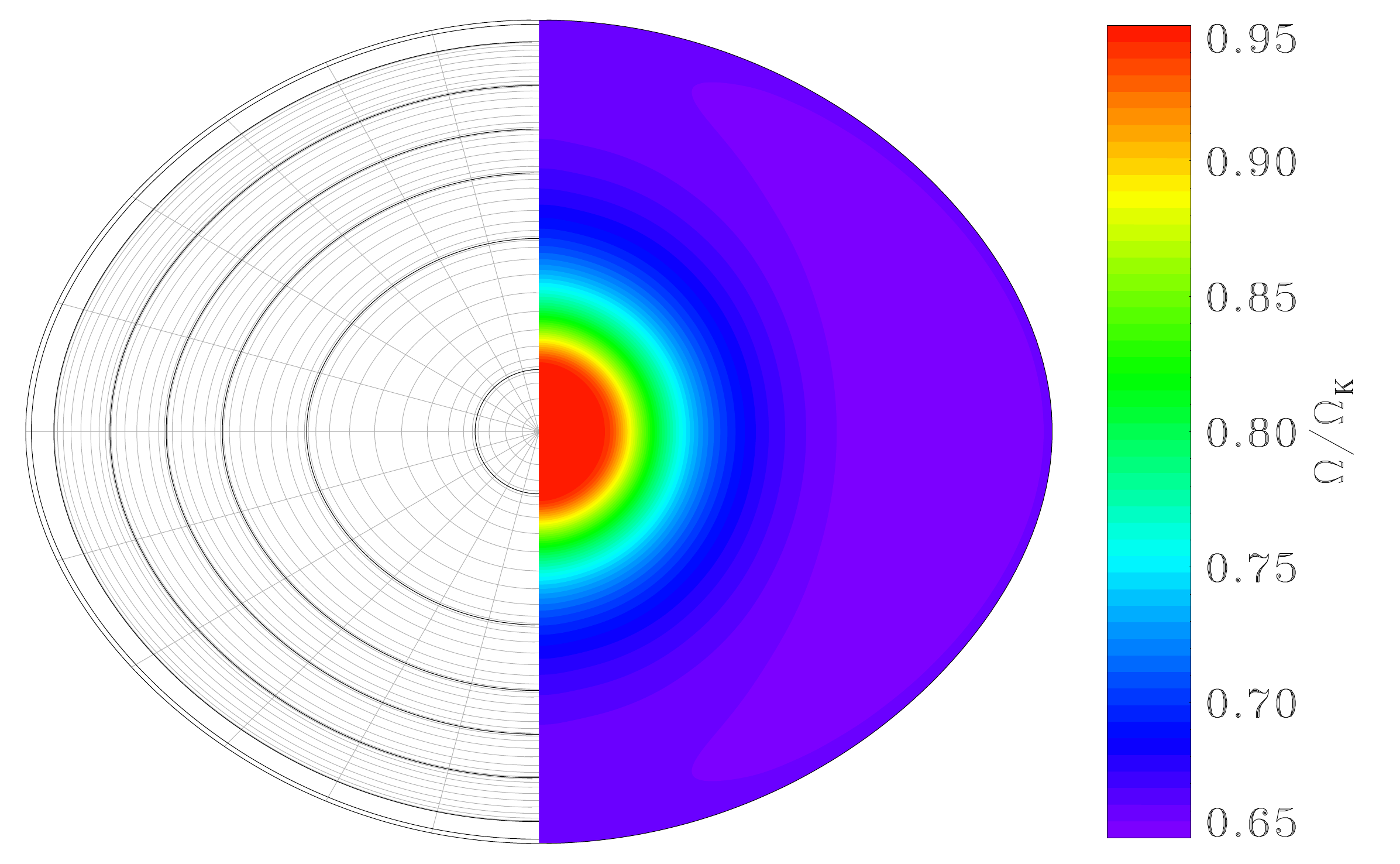}
\caption{Multi-domain coordinate system and 2D rotation profile
in an ESTER model.}
\label{fig:ESTER}
\end{figure}

Models from the ESTER project fully take into account the effects of rotation.
These include: centrifugal deformation, gravity darkening, and baroclinic
effects which result from the mismatch between isobars and isotherms (given that
the temperature profile is determined via the energy conservation equation). 
These effects lead to meridional circulation as well as differential rotation
which depends both on $r$ and $\theta$, as illustrated in Fig.~\ref{fig:ESTER}.
From a numerical point of view, the ESTER code uses a multi-domain spectral
approach with Chebyshev polynomials in the radial direction, and spherical
harmonics in the horizontal directions.  The multi-domain approach is
illustrated in Fig.~\ref{fig:ESTER}.

\subsection{Non-adiabatic pulsations}

The set of non-adiabatic pulsation equations implemented in the TOP code are the
perturbed continuity Equation, Euler's equation, Poisson's equation, the
perturbed energy conservation equation, and the perturbed energy flux equation:
\begin{eqnarray}
  0 &=& \frac{\delta\rho}{\rho_0} + \div \vect{\xi} \\
  0 &=& \vlp^2 \vect{\xi} - 2i \vect{\Omega} \times \vlp \vect{\xi} \nonumber \\
    &&- \vect{\Omega} \times \left( \vect{\Omega} \times \vect{\xi} \right)
     -  \vect{\xi} \cdot \grad \left( \varpi \Omega^2 \vect{e}_{\varpi}\right) 
     -  \frac{P_0}{\rho_0} \grad \left(\frac{\delta P}{P_0}\right) \nonumber \\
    &&+  \frac{\grad P_0}{\rho_0} \left(\frac{\delta\rho}{\rho_0}
     -  \frac{\delta P}{P_0}\right)
     -  \grad \Psi
     +  \grad \left(\frac{\vect{\xi}\cdot\grad P_0}{\rho_0}\right)  \nonumber \\
    &&+ \frac{\left(\vect{\xi}\cdot\grad   P_0\right)\grad\rho_0 -
        \left(\vect{\xi}\cdot\grad\rho_0\right)\grad P_0}{\rho_0^2} \\
  0 &=& \lapl \Psi - 4 \pi G \left(\rho_0\frac{\delta\rho}{\rho_0} 
     - \vect{\xi}\cdot\grad\rho_0\right) \\
  0 &=& - i\vlp \rho_0 T_0 \delta S
     +  \epsilon_0 \rho_0 \left( \frac{\delta \epsilon}{\epsilon_0}
     +  \frac{\delta \rho}{\rho_0} \right)  \nonumber \\
    &&- \div \delta \vect{F}
     +  \vect{\xi} \cdot \grad \left(\div \vect{F}_0 \right)
     -  \div \left[ \left( \vect{\xi} \cdot \grad \right) \vect{F}_0 \right] \\
\delta\vFR &=& \left[ \left(1 + \chi_T\right) \frac{\delta T}{T_0}
     +  \chi_{\rho} \frac{\delta \rho}{\rho_0} \right]\vFR_0
     - \chi_0 \Biggl[ T_0 \grad \left(\frac{\delta T}{T_0}\right) \nonumber \\
    &&+ \vect\xi \cdot \grad \left( \grad T_0 \right)
     -  \grad \left( \vect{\xi} \cdot\grad T_0\right) \Biggr]
\end{eqnarray}
where $\rho$ is the density, $\vect{\xi}$ the Lagrangian displacement, $\omega$
the pulsation frequency, $m$ the azimuthal order, $\Omega$ the rotation profile
(which depends on $r$ and $\theta$), $\vect{\Omega}$ the rotation vector (\ie\
$\Omega\vect{e}_z$), $\varpi$ the distance from the rotation axis, $P$ the
pressure, $\Psi$ the gravitational potential, $T$ the temperature, $S$ the
entropy, $\epsilon$ the nuclear energy generation rate, $\vect{F}$ the energy
flux, $\vFR$ the radiative energy flux, $\chi$ the opacity, $\chi_T$ the
logarithmic derivative of $\chi$ with respect to $T$, \ie\ $\dpart{\chi}{T}$,
and $\chi_{\rho}$ the logarithmic derivative of $\chi$ with respect to $\rho$. 
Quantities with the subscript ``$0$'' correspond to background equilibrium
quantities whereas those preceded by a $\delta$ are  Lagrangian perturbations. 
We note that the last term in Euler's equation is specific to non-barotropic
models. In addition to the above equations, perturbed equations of state and
opacities must be included, as well as a variety of boundary conditions which
ensure regularity of the solution in the centre, a non-divergent gravitational
potential perturbation at infinity, and the appropriate behaviour of the 
pressure and temperature perturbations at the surface.  Overall, the system
comprises 10 equations with 10 unknowns.  The numerical cost in solving this
system is quite high as shown in Table~\ref{tab:numerical_cost} which gives the
computational times and RAM for various radial and horizontal resolutions for
given numbers of processors.

\begin{table}
\centering
\caption{Numerical cost for calculating non-adiabatic modes.}
\label{tab:numerical_cost}
  \begin{tabular}{ccccc}
  \hline
  $\mathbf{N}_{\mathrm{r}}$ & $\mathbf{N}_{\mathrm{h}}$ &
  \textbf{RAM (Gb)} & \textbf{Time (min)} & \textbf{Num. proc.} \\
  \hline
  400 & 10 & 3.5  &    &    \\
  400 & 15 & 7.9  &    &    \\
  400 & 20 & 13.4 &  5 &  4 \\
  400 & 29 & 28.0 & 10 &  8 \\
  400 & 40 & 52.7 & 22 &  8 \\
  400 & 50 & 82.3 & 26 & 16 \\
  \hline
  \end{tabular}
\end{table}

Using a similar approach as in the adiabatic case, it is possible to derive an
integral expression for the frequency (but which will not benefit from a
variational principle), as well as for the damping rate.  The latter is known as
the work integral and is used to increase the accuracy with which the damping
rate is calculated.  The relative difference between the numerical and integral
values of the frequency is of the order of $10^{-4}$, whereas it is around
$10^{-2}$ to $10^{-1}$ for the damping rate given its smaller value.

%-----------------------------------------------------------------------
\section{Results}
\label{sect:results}

\subsection{Frequencies and modes}

\begin{figure}[h]
\centering
\includegraphics[width=\hsize,clip]{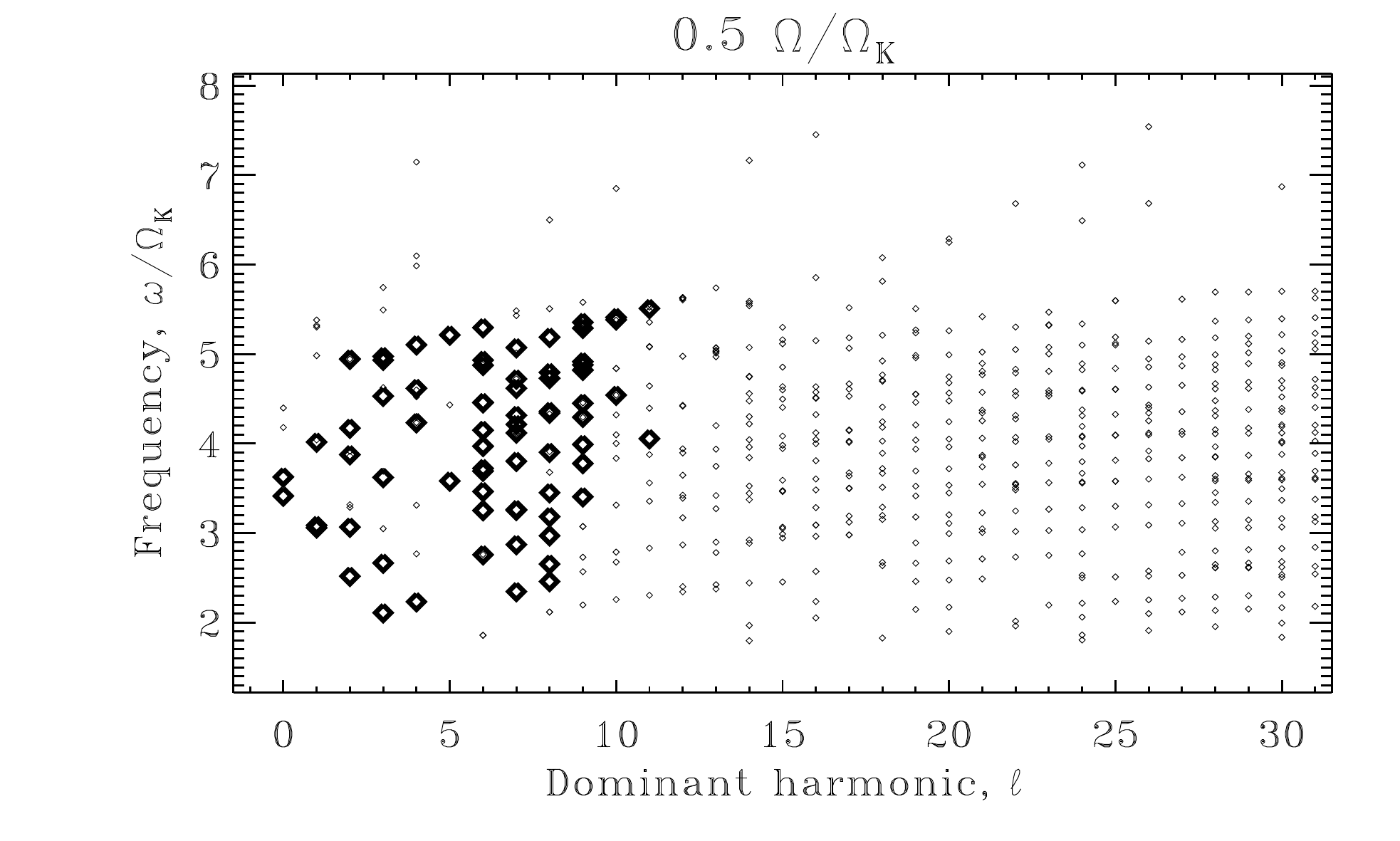}
\caption{Frequency spectrum as a function of dominant spherical harmonic.
The hollow diamonds correspond to unstable modes.}
\label{fig:frequencies}
\end{figure}

\begin{figure}[h]
\centering
\includegraphics[width=\hsize,clip]{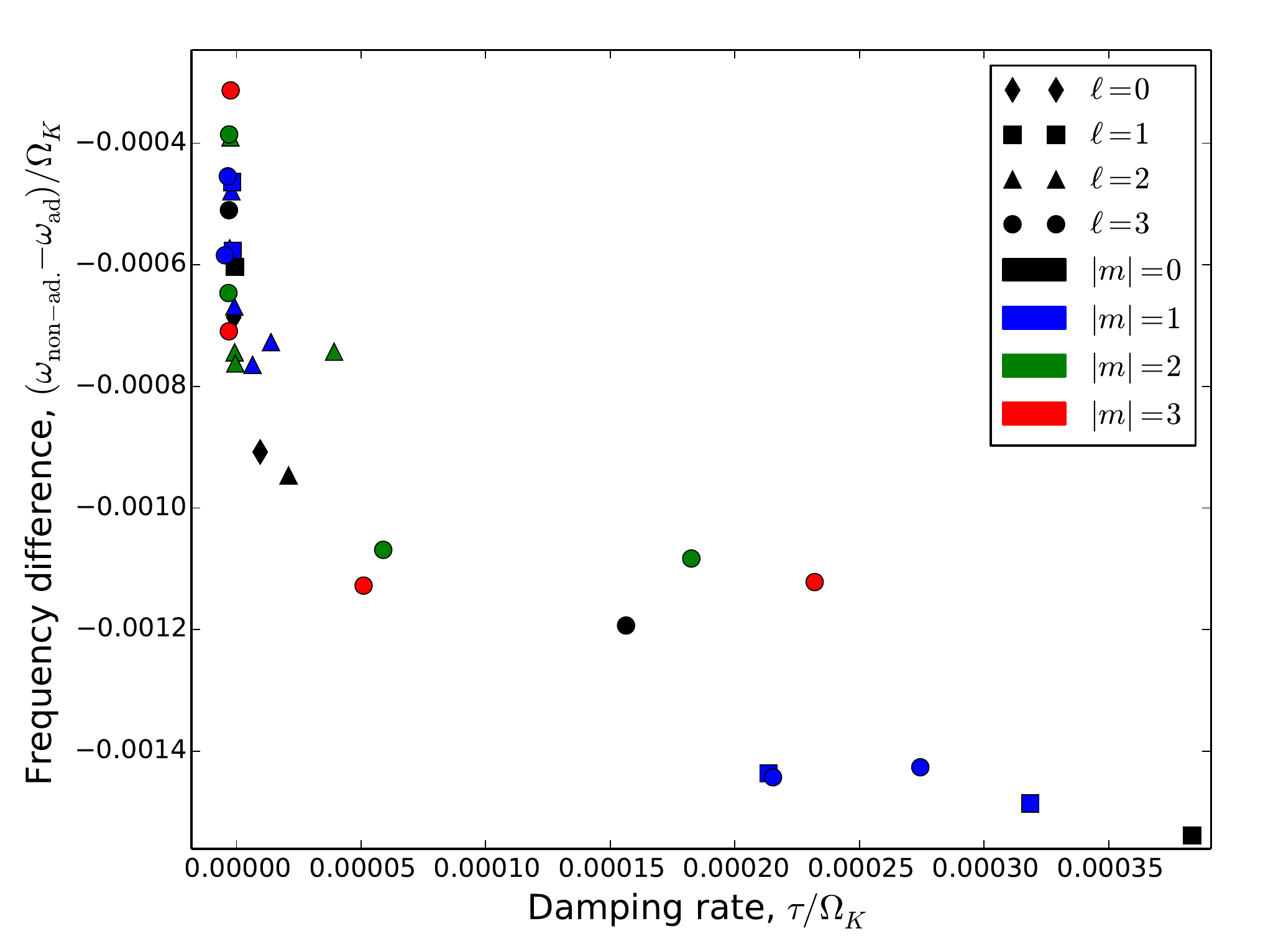}
\caption{Difference between adiabatic and non-adiabatic frequencies
as a function of the damping rate.}
\label{fig:ad_vs_nonad}
\end{figure}

Figure~\ref{fig:frequencies} shows a typical frequency spectrum at
$0.5\,\OmegaK$.  The hollow diamonds correspond to modes which are excited via
the $\kappa$-mechanism.  For the same model, Fig.~\ref{fig:ad_vs_nonad} shows
the difference between adiabatic and non-adiabatic frequencies as a function of
the damping rate.  As expected, the largest frequency differences occur
for the largest damping rates, \ie\ where non-adiabatic effects are the
strongest.

\begin{figure}[h]
\centering
\includegraphics[width=\hsize,clip]{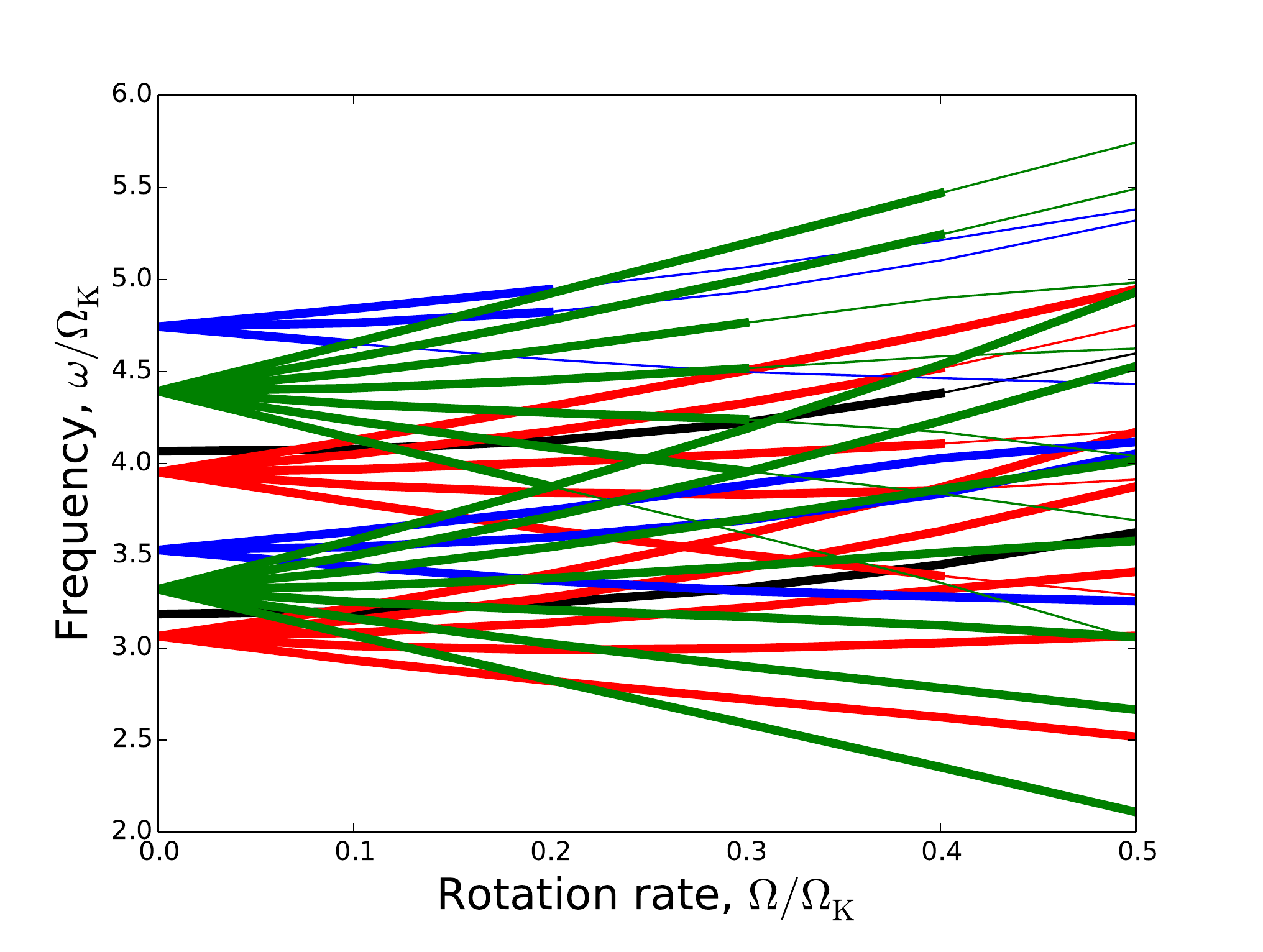}
\caption{Frequencies multiplets as a function of the rotation rate.  The thick
lines corresponds to excited modes.}
\label{fig:multiplets}
\end{figure}

One can look more specifically at frequency multiplets.  In
Fig.~\ref{fig:multiplets}, we show the evolution of various frequency multiplets
as a function of the rotation rate.  The thick lines correspond to excited
modes.  As can be seen, the prograde modes tend to remain excited longer, in
agreement with the results in \cite{Lee2008}.  A look at the work integrals
shows that modes are being stabilised by the very near surface layers, starting
with the retrograde modes (see Fig.~2 of \cite{Reese2015}).

\begin{figure*}[htbp]
\centering
\includegraphics[width=0.23\hsize,clip]{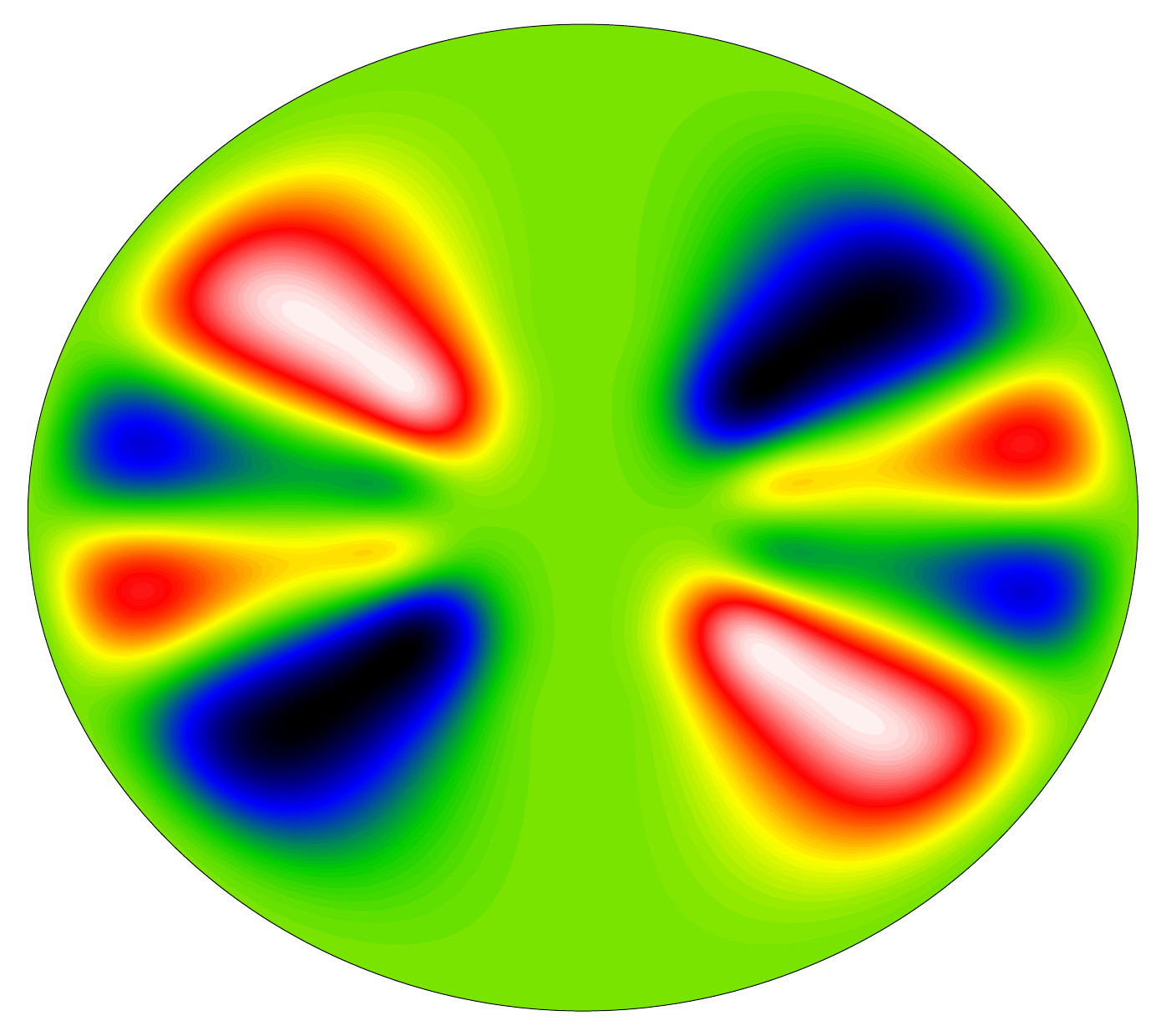} \hfill
\includegraphics[width=0.23\hsize,clip]{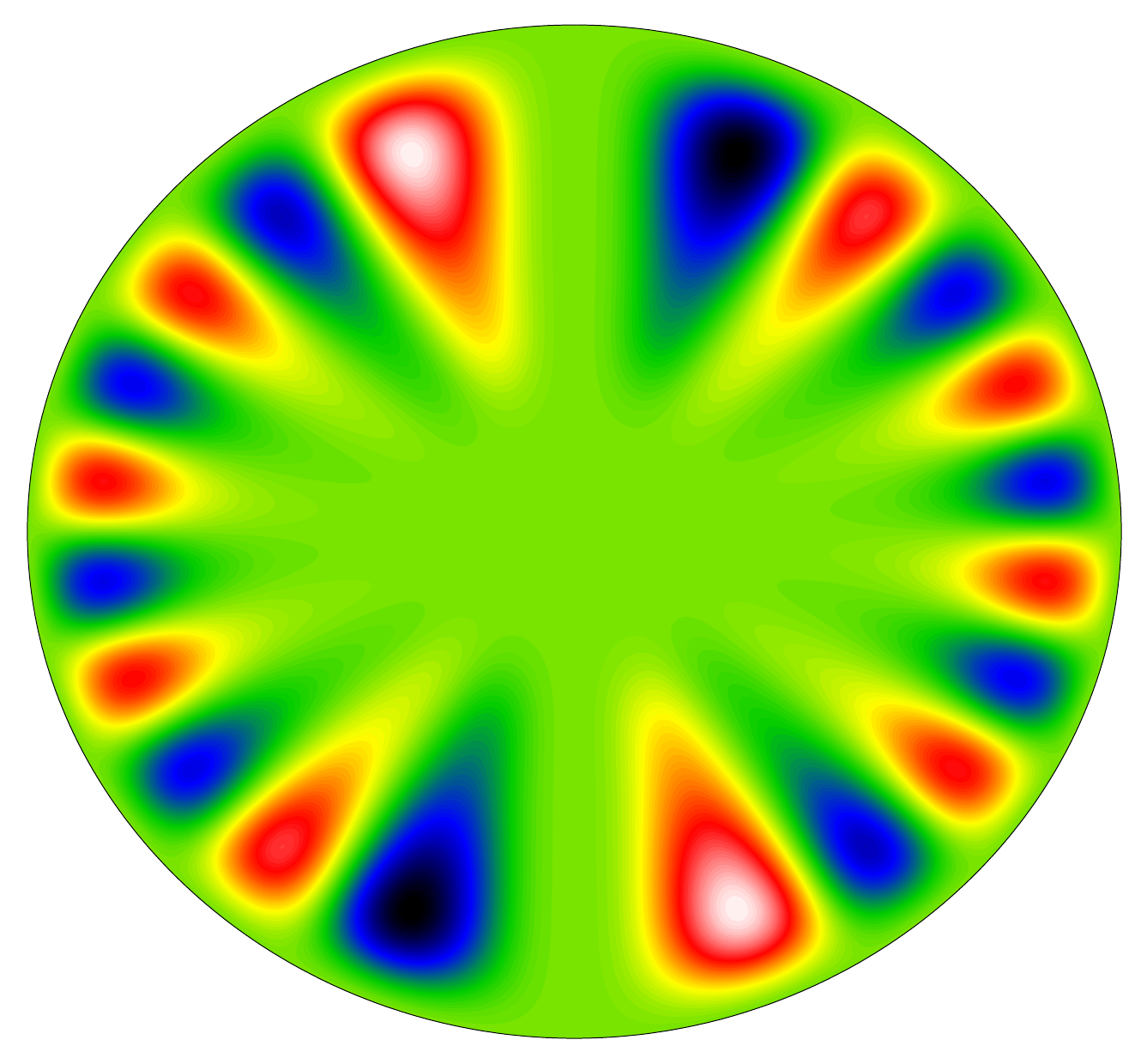} \hfill
\includegraphics[width=0.23\hsize,clip]{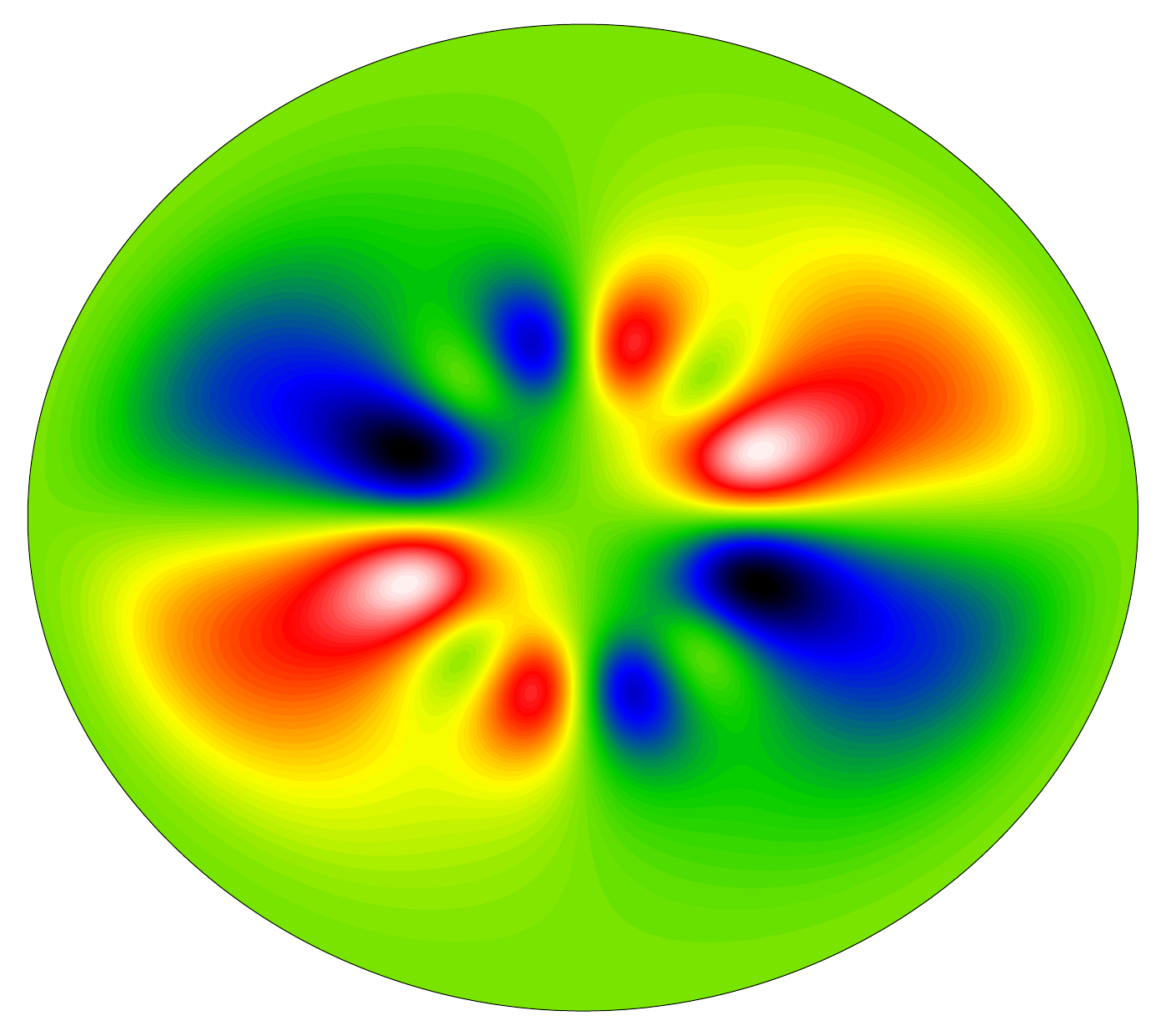} \hfill
\includegraphics[width=0.23\hsize,clip]{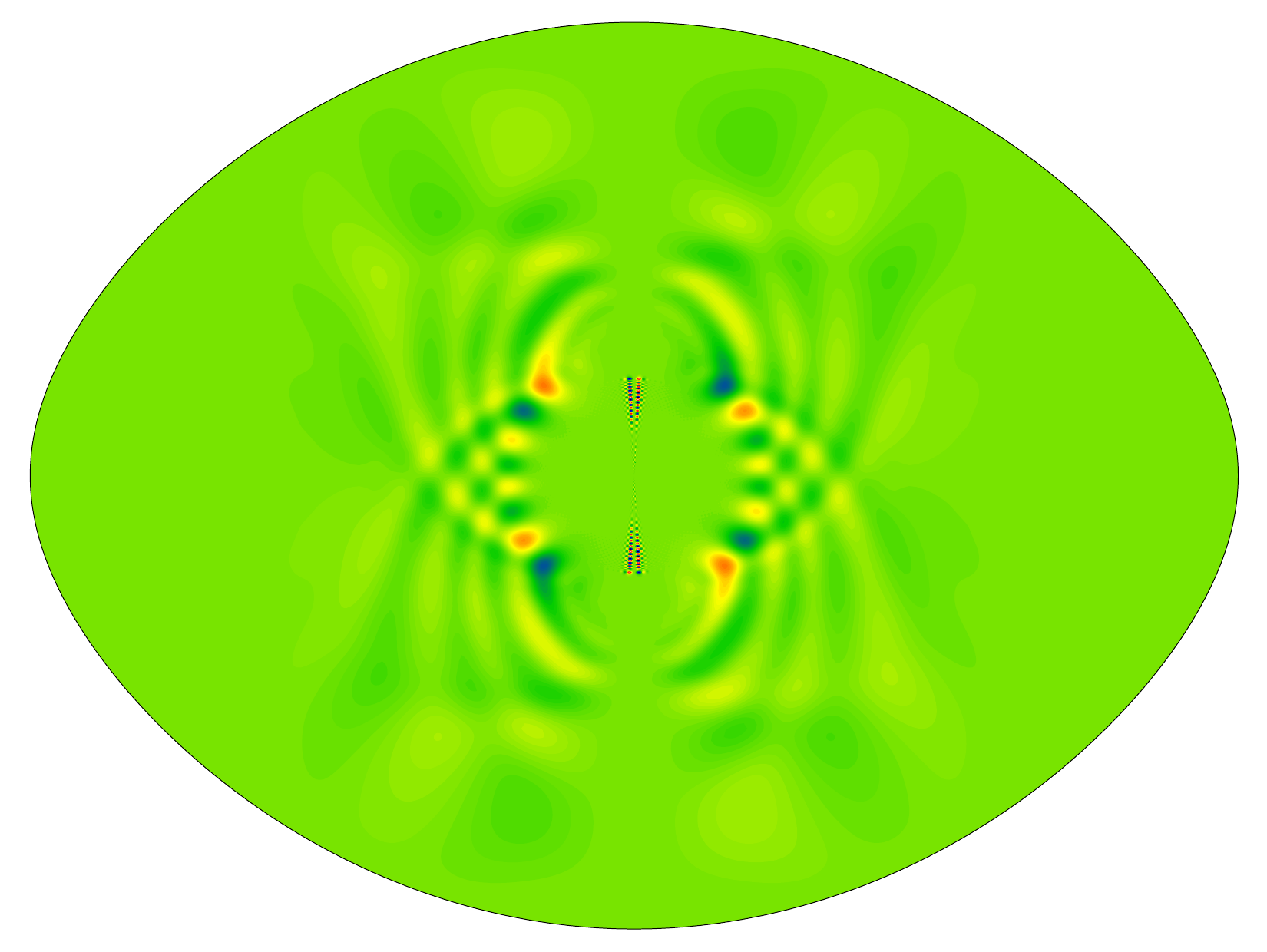}
\caption{Different modes which are excited.  These are, from left to right:  an
island mode, a whispering gallery mode, a mixed mode, and a rosette mode.}
\label{fig:excited_modes}
\end{figure*}

Different types of modes are excited.  These include island modes and chaotic
modes for the acoustic end of the spectrum \cite{Lignieres2009}, as well as
mixed modes \cite{Ouazzani2015} and rosette modes \cite{Ballot2012}, as
illustrated in Fig.~\ref{fig:excited_modes}.

\subsection{Amplitude ratios}

\begin{figure}[h]
\centering
\includegraphics[width=\hsize,clip]{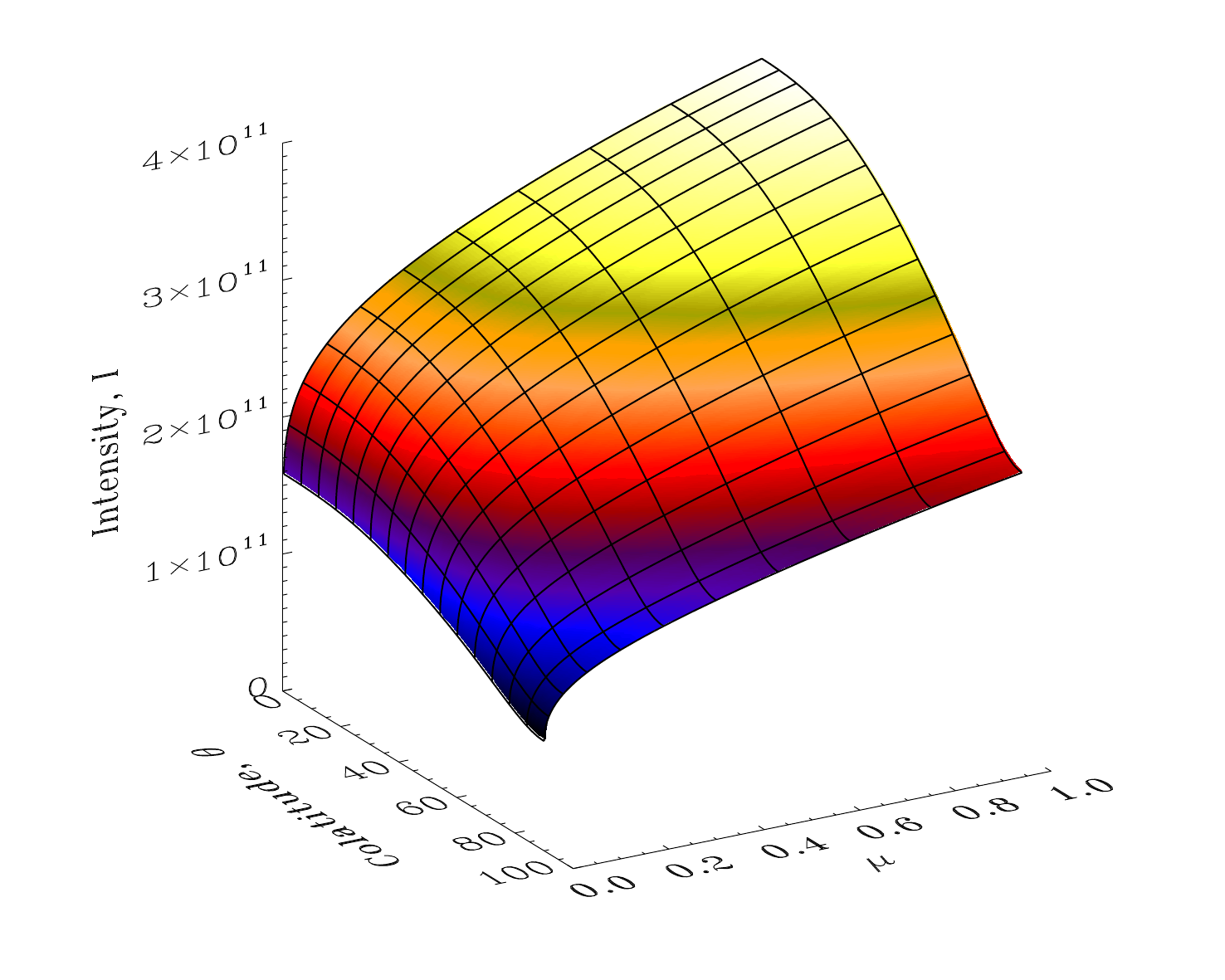}
\caption{Intensity in the Str{\"o}mgren photometric band as
a function of the colatitude and $\mu$ for a model rotating at
$0.8\,\OmegaK$.}
\label{fig:stromgren_u}
\end{figure}

Amplitude ratios were calculated for Str{\"o}mgren and Johnson-Cousins
photometric bands, and in the bolometric case using the formalism
given in \cite{Reese2013}.  In order
to calculate the intensity as a function of effective temperature,
effective gravity, and orientation of the surface, we applied the
following approximate formula:
\begin{equation}
I(\Teff,\geff,\mu) = I_0(\Teff) h(\mu,\Teff,\geff)
\end{equation}
where $I_0(\Teff)$ was obtained from the integral of a blackbody spectrum
weighted by the various photometric filter profiles,
and $h(\mu,\Teff,\geff)$ was
obtained from \cite{Claret2000}, and $\mu = \cos\varphi$ where $\varphi$ is the
angle between the line-of-sight and the normal to the  surface. 
Figure~\ref{fig:stromgren_u} shows the resultant function obtained for the
Str{\"o}mgren u filter for a star at $0.8\,\OmegaK$.  The decrease as a function
of colatitude corresponds to gravity darkening, whereas the increase as a
function of $\mu$ values is caused by limb darkening.

\begin{figure}[h]
\centering
\includegraphics[width=\hsize,clip]{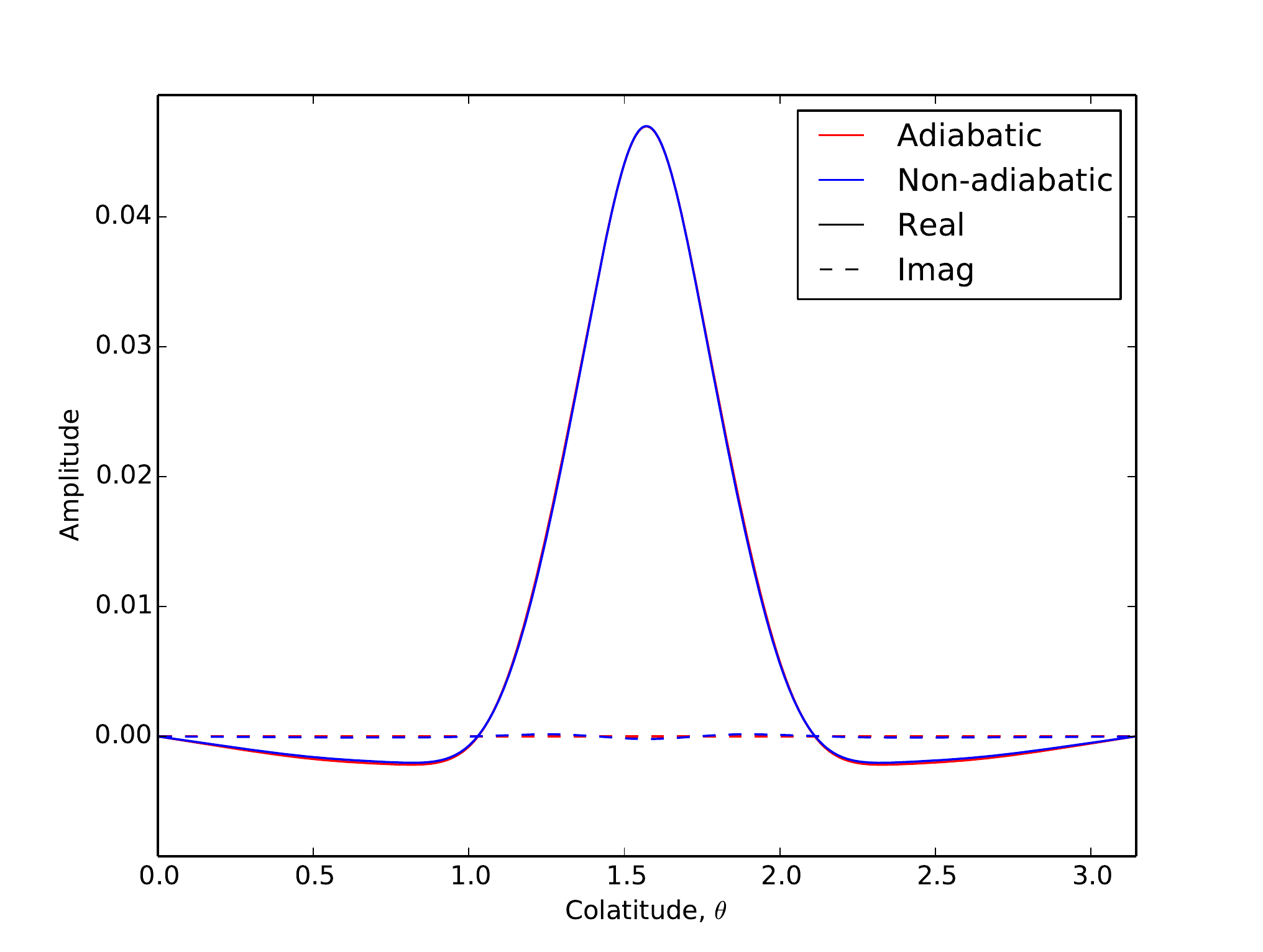} \\
\includegraphics[width=\hsize,clip]{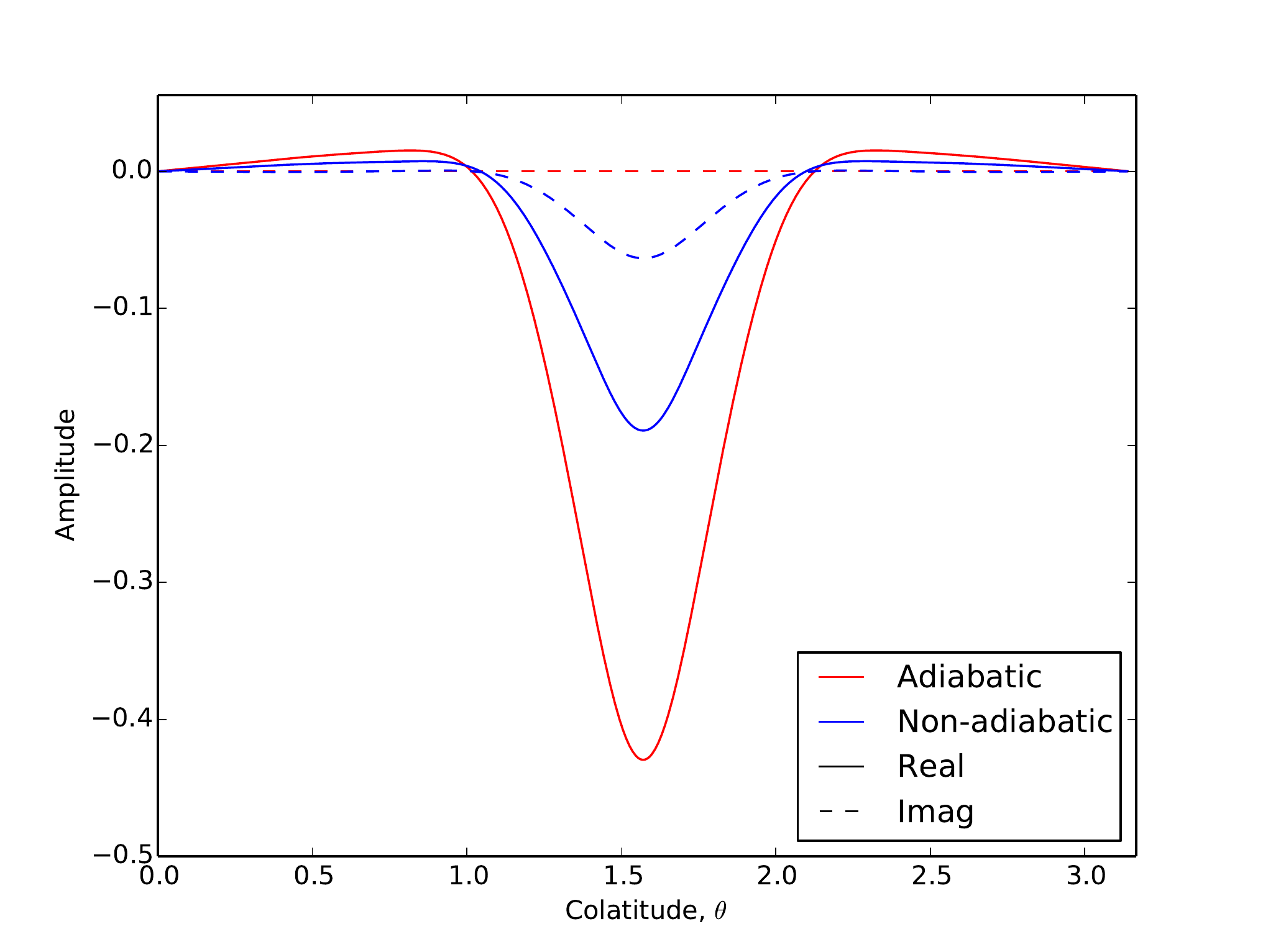}
\caption{Surface vertical Lagrangian displacement (top panel)
and surface $\dTeff/\Teff$ profiles (lower panel) for a given mode
calculated with and without the adiabatic approximation.}
\label{fig:surface_profiles}
\end{figure}

Figure~\ref{fig:surface_profiles} shows the surface Lagrangian displacement, $\xi_r$, and surface $\dTeff/\Teff$ profiles
for a given pulsation mode calculated with and without the adiabatic
approximation. In contrast to $\xi_r$, the $\dTeff/\Teff$ profiles show a
non-negligible imaginary part in the non-adiabatic case.  This corresponds
to a phase shift between the displacement and temperature variations, as
expected in such a case.  As also anticipated, the modulus of
the effective temperature variations, $|\dTeff|$, is smaller in the
non-adiabatic case.  These illustrate the importance of taking
non-adiabatic effects into account prior to calculate mode visibilities,
amplitude ratios, and phase differences.

\begin{figure}[h]
\centering
\includegraphics[width=\hsize,clip]{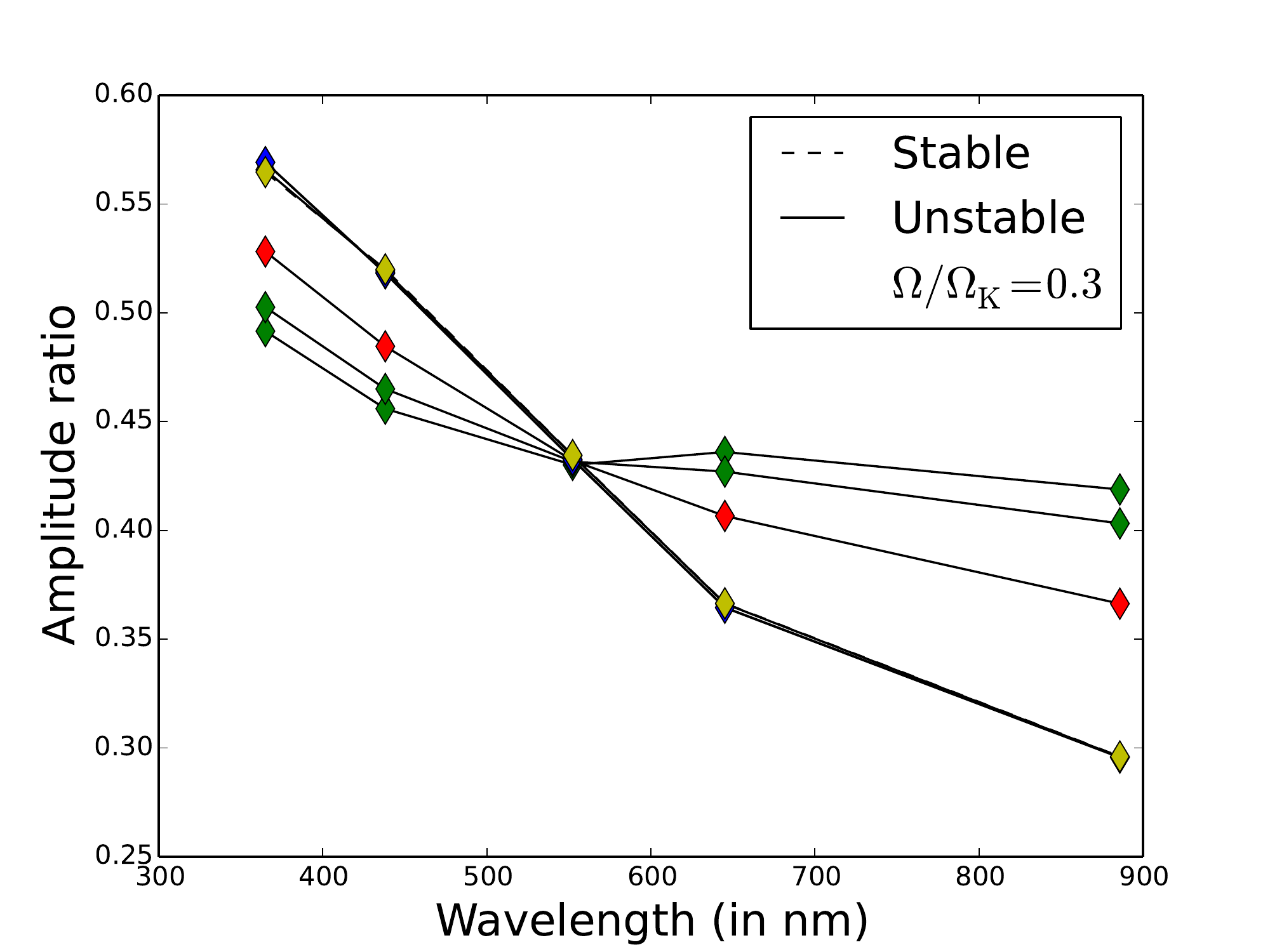} \\
\includegraphics[width=\hsize,clip]{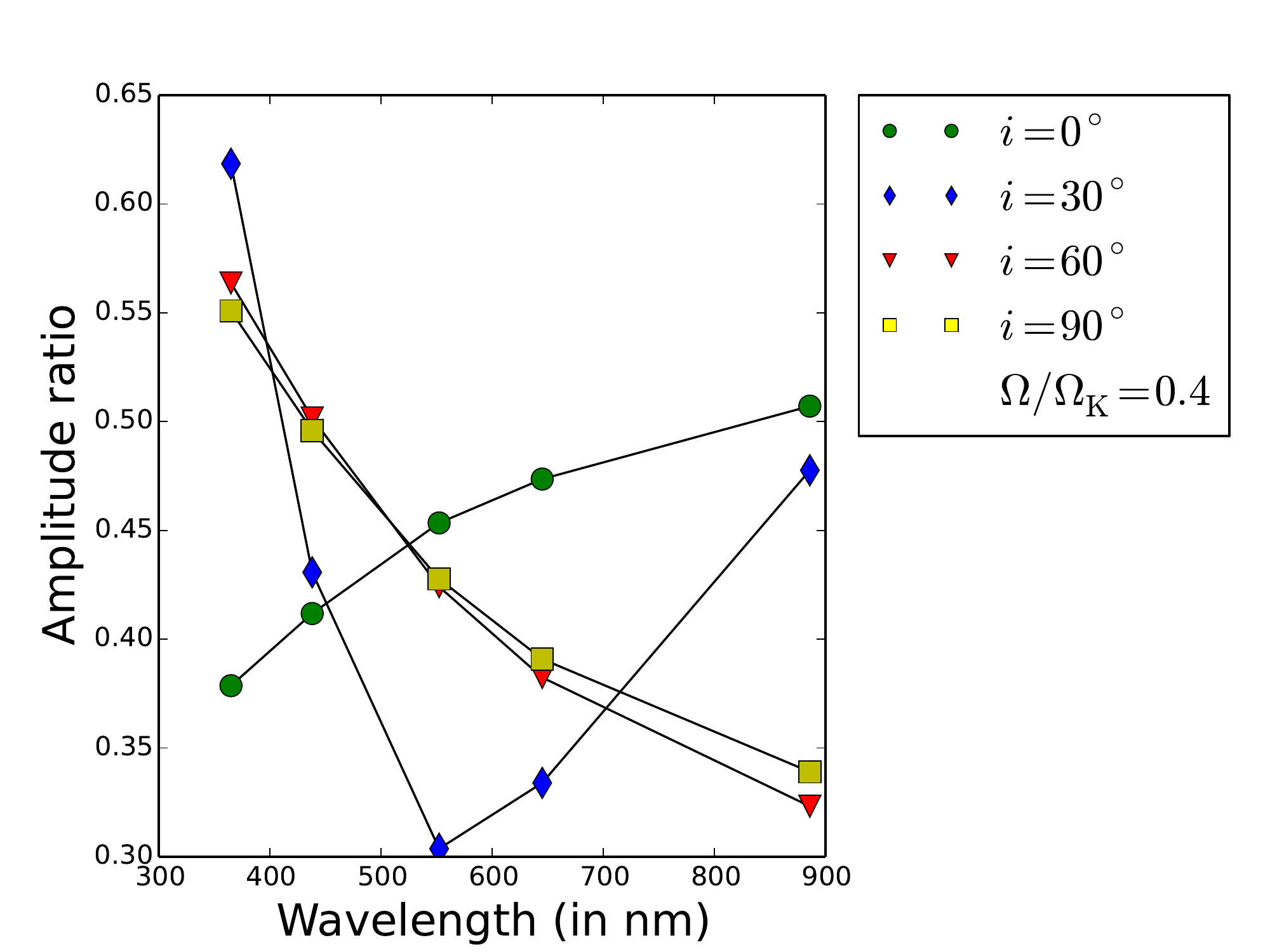}
\caption{Amplitude ratios for an $\l=3$ multiplet at an inclination
of $i=30^{\circ}$ (upper panel)
and for an $(\l,\,m)=(2,0)$ mode at 4 different inclinations
(lower panel).}
\label{fig:multiplet_amplitude_ratios}
\end{figure}

Figure~\ref{fig:multiplet_amplitude_ratios} shows a set of amplitude
ratios for a given multiplet for a model with an inclination of $i=30^{\circ}$,
and for a given mode at 4 different inclinations.  In full agreement with
previous results \cite{Daszynska_Daszkiewicz2002, Townsend2003b}, amplitude
ratios become dependant both on the azimuthal order, $m$, and on the
inclination as rotation is introduced.  This, of course, complicates mode
identification using amplitude ratios although we do note
that \cite{Reese2013b} proposed an alternate mode identification
strategy in such stars.

\subsection{Line profile variations (LPVs)}

Another way of constraining mode identification is through LPVs.  In
the present case, we reuse some of the formulas from \cite{Reese2013}
with an additional Doppler term from both the stellar rotation and
the oscillatory movements.  The ray profile is modelled in crude way
using a blackbody spectrum (thereby accounting for gravity darkening)
as a background and a Gaussian profile
as an intrinsic absorption ray.  Furthermore, the contributions of
$\dTeff$ and $\dgeff$ are neglected as are any dependence of the 
absorption ray on $\Teff$ and $\geff$.  A rudimentary description
of limb-darkening is also included.

\begin{figure*}[h]
\centering
\includegraphics[width=0.75\hsize,clip]{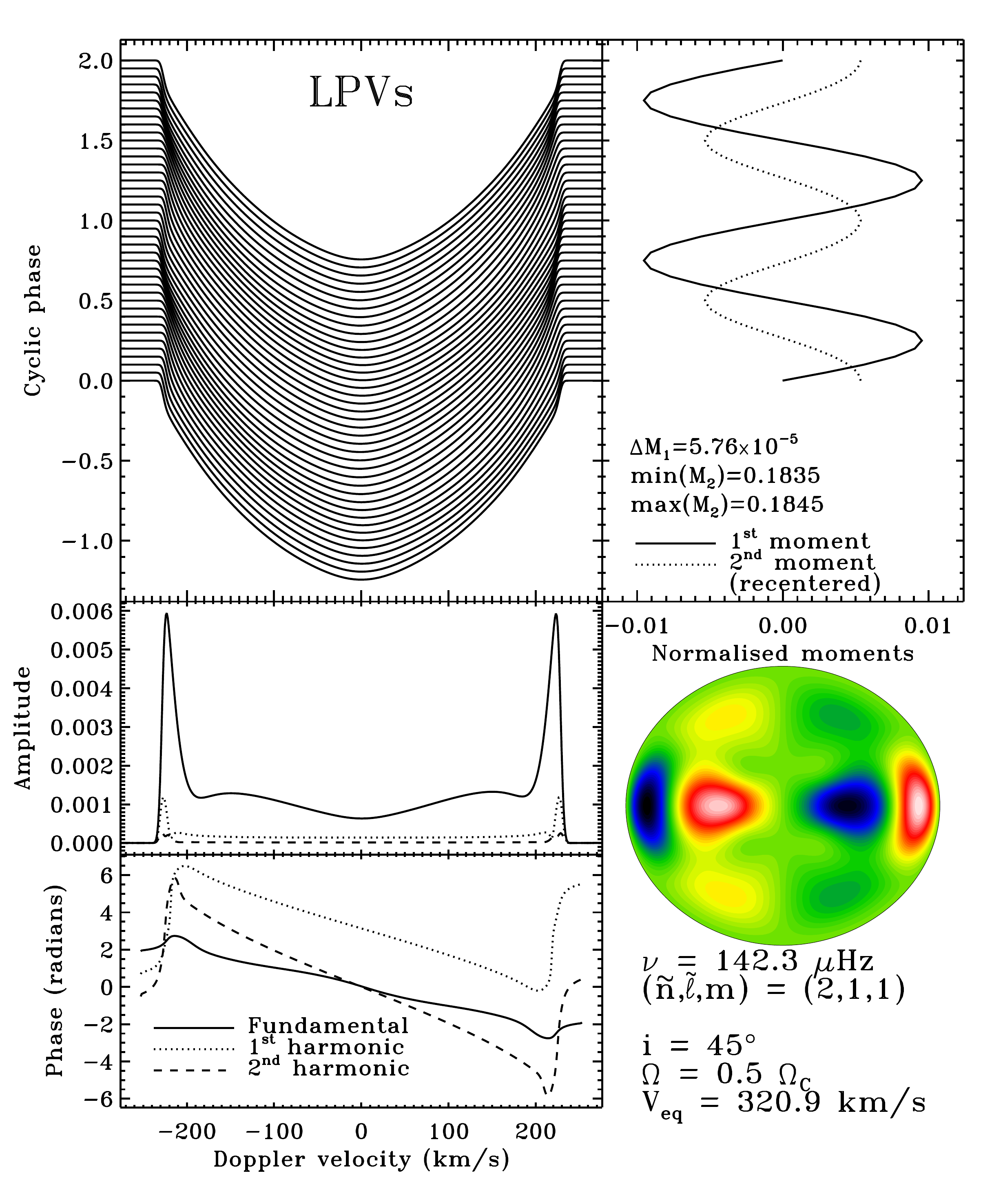}
\caption{Line profile variations for an $(n,\,\l,\,m)=(1,3,1)$
mode (top left panel), first and second moments (top right panel),
amplitudes and phases of the first three
harmonics of the variations across the line profile (lower left panels),
and meridional cross-section of the mode (lower right panel).}
\label{fig:lpvs1}
\end{figure*}

\begin{figure*}[h]
\centering
\includegraphics[width=0.75\hsize,clip]{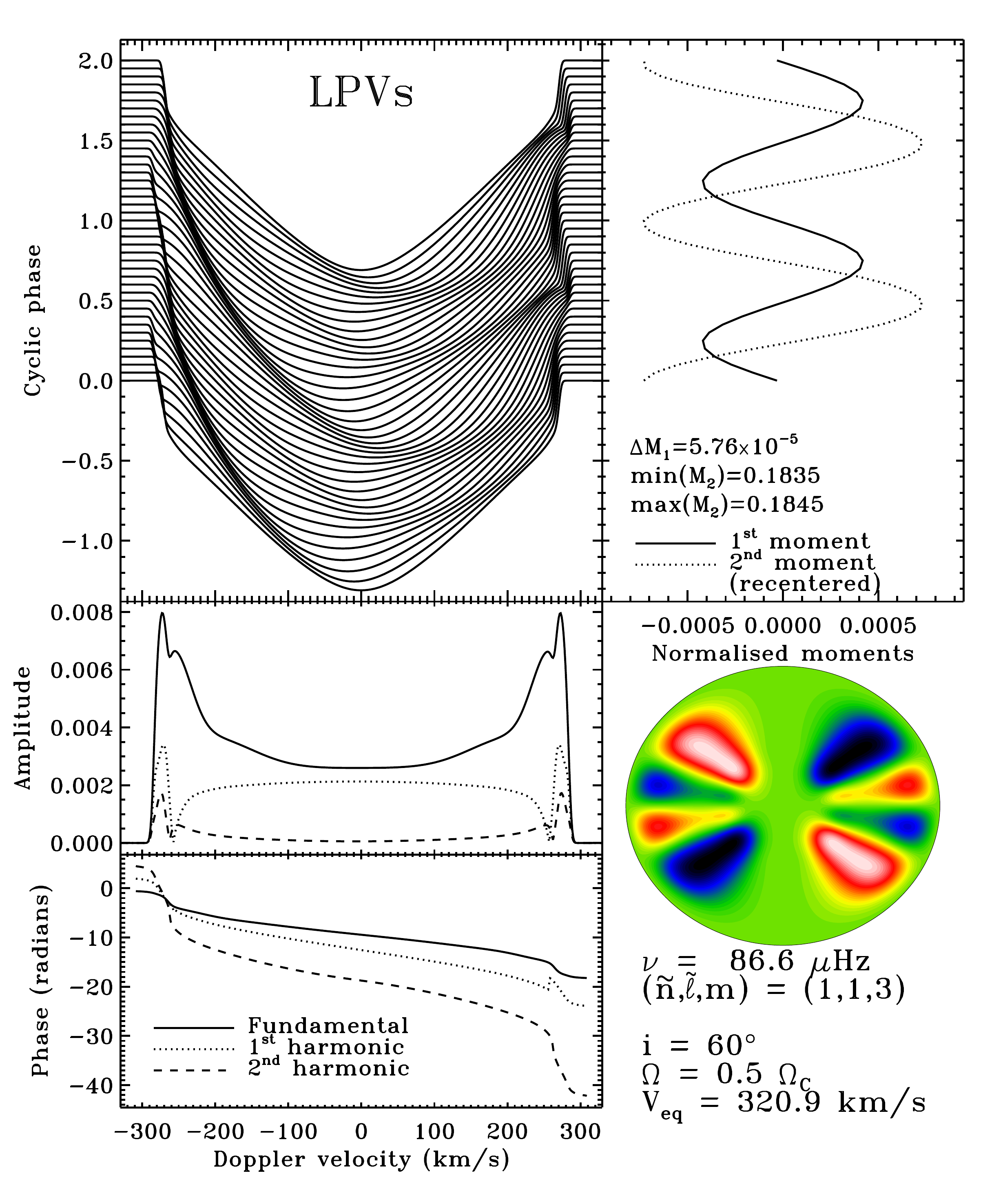}
\caption{Same as Fig.~\ref{fig:lpvs1} but for the
$(n,\,\l,\,m)=(0,6,3)$ mode.}
\label{fig:lpvs2}
\end{figure*}

Figures~\ref{fig:lpvs1} and~\ref{fig:lpvs2} show two examples of
LPVs calculated in a
model rotating at $0.5\,\OmegaK$.  These plots also contain the
first and second moments, the amplitudes and phases of the first three
harmonics of
the variations for a particular Doppler position, and a meridional
cross-section of the modes.  A note worthy feature is that most
of the variations take place in the wings of the absorption ray,
which is different than what takes place in the non-rotating case.
This feature is fairly frequent at rapid rotations and has been
suggested as a possible explanation for macro-turbulence in
massive stars (Aerts, private communication).

%-----------------------------------------------------------------------
\section{Conclusions}

The ability to carry out non-adiabatic pulsation calculations in rapidly rotating stars
is an important step forward in being able to understand and interpret
seismic data for such stars.  Indeed, it is now possible to predict
which modes are excited, and it will lead to more realistic predictions
for amplitude ratios, phase differences, and line profile variations.
The next steps forward include: gaining a better understanding of how
rotation interacts with non-adiabatic effects, understanding what causes
the differences between prograde and retrograde modes, including a
realistic atmosphere in the stellar model when calculating amplitude ratios and
line profile variations, and identifying modes in observed spectra
using multi-colour photometry and spectroscopy.

\section*{Acknowledgements}
DRR acknowledges financial support from the ``Programme National de Physique
Stellaire'' (PNPS) of CNRS/INSU, France.
%-----------------------------------------------------------------------
\bibliography{biblio}

\end{document}